\documentclass[draft]{agujournal}
\draftfalse

\journalname{Geophysical Research Letters}

\begin{document}

\title{The Sensitivity of Euro-Atlantic Regimes to Model Horizontal Resolution}

%
%

\authors{K. Strommen\affil{1}, I. Mavilia\affil{2}, S. Corti\affil{2}, M. Matsueda\affil{3,1}, P. Davini\affil{2}, J. von Hadenberg\affil{2}, P-L. Vidale\affil{4}, R. Mizuta\affil{5}}

\affiliation{1}{Department of Physics, University of Oxford, Oxford, United Kingdom}
\affiliation{2}{Istitutio di Scienze dell'Atmosfera e del Clima, Consiglio Nazionale delle Richerche, Italy}
\affiliation{3}{Center for Computational Sciences, University of Tsukuba, Tsukuba, Japan}
\affiliation{4}{Department of Meteorology, University of Reading, Reading, United Kingdom}
\affiliation{5}{Meteorological Research Institute, Tsukuba, Japan}

\correspondingauthor{K. Strommen}{Oxford University, Clarendon Laboratory, AOPP, Parks Road, OX1 3PU, Oxford. E-mail: kristian.strommen@physics.ox.ac.uk}

\begin{keypoints}
\item Climate models have difficulty representing North Atlantic regime structure correctly.
\item Increasing horizontal resolution improves the significance of regime clustering across multiple models.
\item Spatial patterns and persistence levels of regimes do not necessarily improve with increased resolution.
\end{keypoints}

%
%

\begin{abstract}
There is growing evidence that the atmospheric dynamics of the Euro-Atlantic sector during winter is driven in part by the presence of quasi-persistent regimes. However, general circulation models typically struggle to simulate these, with e.g. an overly weakly persistent blocking regime. Previous studies have showed that increased horizontal resolution can improve the regime structure of a model, but have so far only considered a single model with only one ensemble member at each resolution, leaving open the possibility that this may be either coincidental or model-dependent. We show that the improvement in regime structure due to increased resolution is robust across multiple models with multiple ensemble members. However, while the high resolution models have notably more tightly clustered data, other aspects of the regimes may not necessarily improve, and are also subject to a large amount of sampling variability that typically requires at least three ensemble members to surmount.
\end{abstract}

\section{Introduction}
\label{sec:introduction}

Predicting the evolution of the atmospheric state over time can be understood as a question of determining likely trajectories along the atmospheres climate attractor in phase space. Over the last two decades, evidence has begun to accumulate that suggests the geometry of this attractor exhibits interesting local structure, which manifests itself in the form of quasi-persistent weather regimes (\citep{Straus2007}, \citep{Straus2010}, \citep{Woollings2010a}, \citep{Woollings2010b}, \citep{Franzke2011}, \citep{Hannachi2017}). In particular, such regimes have been identified in the Euro-Atlantic region, and there is a growing recognition of their importance in modulating European weather (\citep{Ferranti2015}, \citep{Matsueda2018}, \citep{Frame2013}) and, conjecturally, the regional response to anthropogenic forcing (\citep{Palmer1999}, \citep{Corti1999}). Representing these regimes correctly is therefore an important goal for any general circulation model (GCM).

The studies \citep{Dawson2012} and \citep{Dawson2015} demonstrated that a GCMs ability to capture Euro-Atlantic regimes appears to depend on the horizontal resolution of the model. In particular, improvements in the spatial structure, geometric robustness (by which we mean the extent to which the data can be divided into tightly knit clusters), and persistence statistics of the regimes were all identified upon increasing the resolution. However, these studies used only one model, with a single ensemble member at each resolution. This leaves open the question as to how robust this resolution-dependence is across models, as well as the possibility that sampling variability may be playing a role. In this paper, we address these issues by examining the impact of increasing resolution on three different models, each with three ensemble members. Besides examining the impact of resolution, we also evaluate the impact of using an ensemble: by concatenating multiple ensemble members, we can obtain larger datasets, effectively reducing the impact of excessive noise and/or poorly constrained regimes.

We will show, that for all three models considered, the low resolution models struggle to replicate the regime structure seen in re-analysis datasets. None of the nine individual simulations achieve comparable levels of clustering to that of re-analysis, and while the regime patterns on average have a relatively high spatial correlation with those of re-analysis, the spread is often large with some individual members performing notably poorly. Persistence of the blocking regime is also systematically underestimated in all simulations, with the model tending to vacate the regime faster than re-analysis. Increasing the horizontal resolution leads to notably more tightly clustered data, with a few individual high resolution simulations achieving a regime structure comparable to re-analysis. A systematic improvement in the persistence statistics of the blocking regime is also seen across all the models; no such systematic change is identified for the other regimes. This is consistent with the results of the multi-model study conducted in \citep{Schiemann2017}, demonstrating improvements in atmospheric blocking (as measured using more standard European blocking indices) with increased horizontal resolution. However, no systematic improvement in the spatial patterns of the regimes are seen, with the net impact being a slight degradation compared to the low resolution patterns. 

We also show that, for single ensemble members at low resolution, there can be a notable spread around the average value of the metrics in question, suggesting that sampling variability for these quantities can be large. For the low resolution models, one generally needs to use all three members in order to generate regime statistics comparable to re-analysis over the approximately 30-year periods considered, while for high resolution, two ensemble members suffice. This supports the idea that models at lower resolution have too weak regime structure, and that increased resolution can be expected to ameliorate this to some extent. It also highlights the fact that, for models with weaker regime structure, a large sample size of simulation years is necessary to diagnose regimes robustly.

\section{Data and Methods}

\subsection{Data}
\label{sec:data}

We use model data from three models, all run in atmosphere-only mode, covering between 25 and 31 years in the period 1979-2011. Each model produced three simulations at both a `low' and `high' resolution, where the exact meaning of low and high varies between the models. This leaves us with nine low resolution simulations and nine high resolution simulations to compare across: due to the varying nature of the resolution increase, we always group results by model, to see the relative impact in each model. We also note that the monikers `low' and `high' resolution are essentially arbitrary here, and are used simply for convenience to denote the lower/higher of the two available resolutions, rather than any objective measure. \added{The study is therefore only concerned with the effect of \emph{increasing} resolution, and not on the exact impact of any \emph{particular} resolution choice.}\explain{Final sentence here added for clarity}

The first model is EC-Earth v3.1, an Earth-system model maintained by the EC-Earth Consortium (\citep{Hazeleger2012}). Its atmospheric component is based on the Integrated Forecasting System (IFS) model cycle 36r4, developed by the European Centre for Medium-Range Weather Forecasts (ECMWF). The integrations were made as part of the Climate SPHINX Project (\citep{Davini2017}), and covered the period 1979-2008. Ten such ensemble members were produced in the SPHINX Project: only 3 were considered in the analysis in order to make the comparison across all models and resolutions as uniform as possible. Results were found to be qualitatively identical irrespective of which 3 ensemble members were selected, so the results presented here used the first three members. The low resolution simulations had a spectral truncation of TL255, or roughly 80km grid-spacing near the equator. The high resolution simulations had a spectral truncation of TL511, corresponding to around 40km grid-spacing. \added{Both use 91 levels in the vertical}. Note that the studies \citep{Dawson2012} and \citep{Dawson2015} considered an earlier version of the same model.

The second model is the UK Met Office model HadGEM3-GA3 (\citep{Walters2011}). The integrations were run as part of the UPSCALE project (\citep{Mizielinski2014}), and cover the period 1986-2011. The low resolution simulations were performed on a N216 grid, corresponding to roughly 60km grid-spacing near the equator, while the high resolution simulations were done on a N512 grid, corresponding to roughly 25km grid-spacing. \added{Both configurations use 85 levels in the vertical.}

Finally, we used the Japanese Meterorological Research Institute (MRI) model AGCM3.2 (\citep{Mizuta2012}). The low resolution version was integrated at TL95 resolution, corresponding to roughly 180km grid-spacing, while the high resolution simulations were integrated at TL319 resolution, corresponding to roughly 60km grid-spacing. \added{Both use 64 levels in the vertical.} The three ensemble members cover the period 1979-2010.

The primary re-analysis product used to act as our reference dataset was the ECMWF dataset ERA-Interim (\citep{Dee2011}), covering the period 1979-2011. To bolster confidence in the results, and to estimate the potential sampling variability of the metrics inherent to the real atmosphere, we also utilized the NCEP/NCAR re-analysis data set (\citep{Kalnay1996}), hereafter referred to as NCEP, \added{and the Japanese 55-year reanalysis (abbr. JRA55: see \citep{Kobayashi2015})}. \explain{The JRA55 dataset was added to address an issue raised by reviewer 2.} In general, the difference between the three data sets for a given metric were small, being an order of magnitude smaller than the model biases and the impacts seen from a resolution change. \added{Because all three datasets show such close agreement, we will only present explicit values for ERA-Interim and NCEP in tables/figures.}

All datasets were first interpolated down to a common 2.5 degree regular grid prior to carrying out computations.

It is important to note that the actual range of resolutions considered in the paper are relatively narrow, leaving open the question as to whether the observed changes could be expected for \emph{any} change in resolution. Also important to note is that none of the high resolution models were tuned separately from the low resolution: the simulations therefore differ only in the resolution itself.

\subsection{Methodology}
\label{sec:methodology}

Regimes are defined using a k-means clustering algorithm, following the method in \citep{Straus2007} (see also \citep{Michelangeli1995}): a regime thereby corresponds to something resembling a fixed point in phase space, around which observed atmospheric states tend to cluster. The algorithm is applied to the daily geopotential height field at 500hPa, considered over a Euro-Atlantic domain defined by $30^{\circ}$-$90^{\circ}$N, $80^{\circ}$W-$40^{\circ}$E. We then restrict the data to the December-January-February winter period for each available year. A climatological cycle is obtained from this field, and smoothed with a 5-day running mean; this smoothed cycle is then removed from the original field to produce a timeseries of daily geopotential height anomalies. In order to make the algorithm tractable, the dimensionality of the field is reduced using an empirical-orthogonal-function (EOF) decomposition. Only the first four (un-normalized) EOFs are retained: these explain more than 50\% of the variance for both models and re-analysis. It was found that using more EOFs, explaining up to 80\% of the variance, produced quantitatively similar results. Restricting to the first four alone also focuses the analysis on the large-scale patterns which we are interested in. The k-means clustering algorithm applied to this final field will then produce clusters that maximize the following \emph{optimal ratio}:
\begin{equation}
\textit{Optimal Ratio} = \frac{\textit{Inter-cluster variance}}{\textit{Intra-cluster variance}},
\end{equation}
where the inter-cluster variance refers to the variance between the cluster centroids (weighted by the number of points in each cluster), and the intra-cluster variance refers to the average variance of the differences between the cluster centroids and the data-points associated to that cluster. A large inter-cluster variance therefore implies that the centroids are well separated from each other, while a small intra-cluster variance implies the points of each cluster are located close to their respective centroid. A large optimal ratio is therefore associated with a more clearly robust regime structure.

The presence of high autocorrelation in the data can influence the k-means clustering algorithm, potentially inflating the optimal ratio. This is exacerbated by the fact that the algorithm will always generate the number of clusters one asks for, meaning large optimal ratios may occur purely by chance. This raises two issues. Firstly, how does one evaluate the statistical significance of the regimes generated? Secondly, given the influence of autocorrelation, which may vary between simulations, how does one compare optimal ratios across multiple models? Both these questions are addressed by defining a `significance' metric in the following manner. A statistical null hypothesis is assumed in which the phase space in question has no particular regime structure, and therefore the atmosphere has no preferred locations or directions of movement in phase space. Concretely, the null hypothesis posits that the phase space is equivalent to that expected from assuming that each of the four coordinates of the atmosphere, in this truncated 4-dimensional phase space, are behaving like independent Markov processes with a fixed mean, variance and lag-1 correlation equaling those of the dataset in question. The assumption of the process being first-order (and therefore using the lag-1 correlation) was justified by plotting the autocorrelation of our datasets and noting that these were very well captured by a basic exponential decay. While skewness in the data can, in principle, also influence clustering, we found little sensitivity to the computed metrics when adding skewness to the null hypothesis, and this was therefore ignored. Randomly generating four such Markov processes defines new atmospheric coordinates which will populate the four-dimensional phase space; by applying the clustering algorithm to this dataset, one computes the optimal ratio for the clusters produced. Repeating this for 500 different synthetic datasets generates a distribution of optimal ratios that could be expected from our null hypothesis. We define the \emph{sharpness} of the clustering to be the percentage of points in this distribution \emph{below} the optimal ratio actually computed with the original dataset. A sharpness close to 100\% therefore implies a large optimal ratio unlikely to have arisen by chance from an atmosphere with no regime structure. Crucially, by effectively `normalizing' the optimal ratio relative to the autocorrelation of the underlying data, the sharpness metric can readily be compared across multiple models. In \citep{Dawson2012}, this metric was referred to simply as `significance', but as we are using this as a metric in and of itself, we rename it to avoid potential confusion.

\begin{figure}[h]
\centering
\includegraphics[width=20pc]{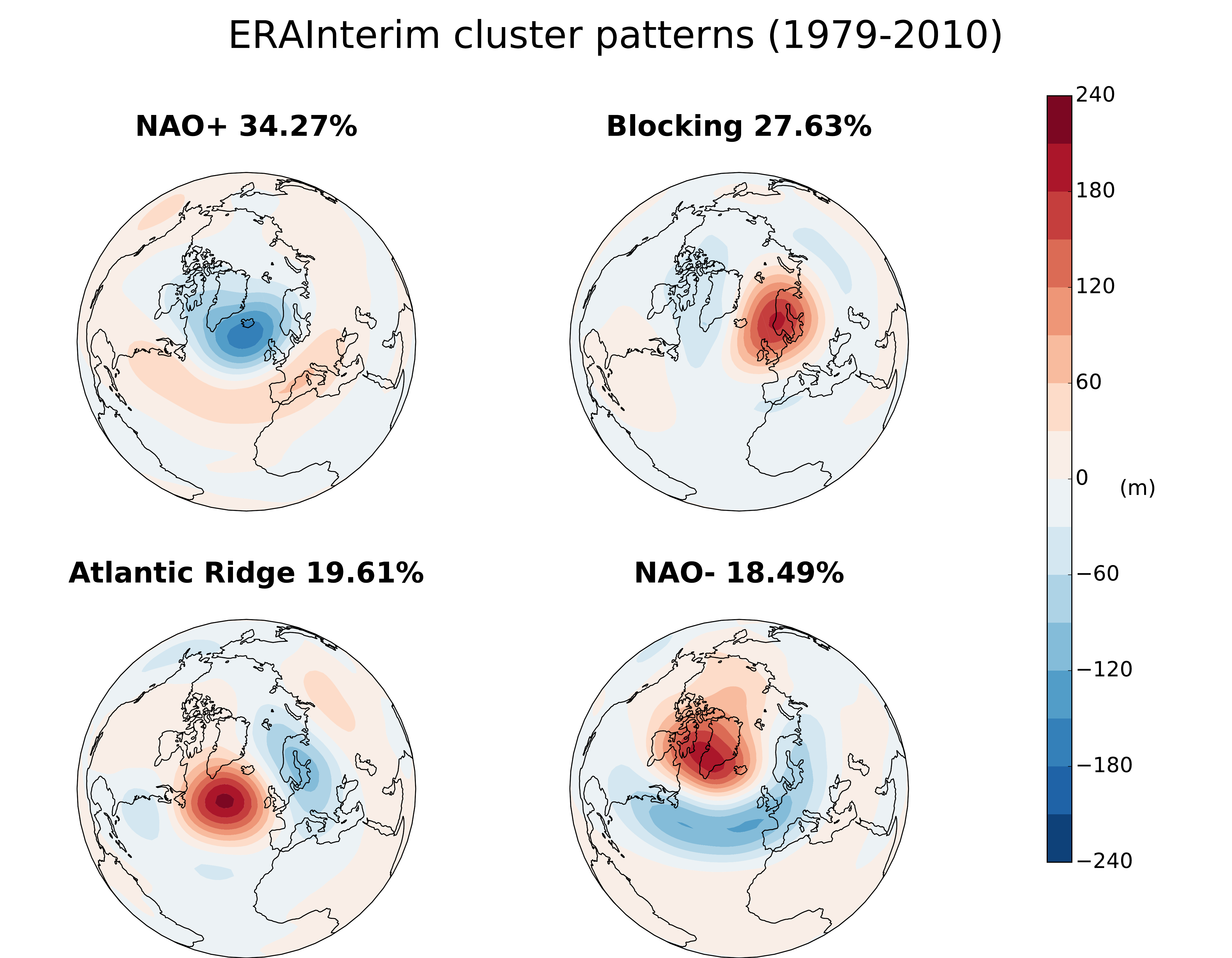}
\caption{Spatial patterns of the four regimes defined by the cluster centroids for ERA-Interim (1979-2010). Obtained by applying k-means clustering to the geopotential height anomalies at 500hPa, restricted to the Euro-Atlantic region. The percentages indicate the frequency of occurrence of that regime during the entire time-period.}
\label{fig:erapatterns}
\end{figure}

Applying the algorithm to re-analysis data shows, as noted in \citep{Dawson2012}, that four clusters are the minimum number required to produce statistically significant regimes (i.e. a sharpness exceeding 95\%). For this reason, we restrict our attention to a four-regime picture, both for re-analysis and our model data. Figure \ref{fig:erapatterns} shows the spatial patterns of these four regimes in the re-analysis data, which agree well with previous studies. These patterns are generated by taking the mean across all the daily fields that are sorted into a given cluster by the algorithm.

We will focus on three metrics for assessing the models representation of regimes. Firstly, the sharpness metric will be used as a measure of the geometric robustness of regimes. A high sharpness suggests strongly defined regimes, but can be obtained with regimes that do not look like those in reality. Therefore, secondly, pattern correlation between the regime patterns of the models and those in reanalysis is used as a measure of how similar to re-analysis the diagnosed regimes are. Finally, we look at the level of day-to-day regime persistence.

\added{Note that EOFs are always computed independently for each dataset in question. In particular, when multiple ensemble members are concatenated, EOFs are computed for the concatenated dataset.}\explain{Added for clarification in response to a question raised by reviewer 1.}

\section{Results}

\subsection{Regime Sharpness}
\label{sec:regime_significance}

\begin{figure}[h]
\centering
\includegraphics[scale=0.32]{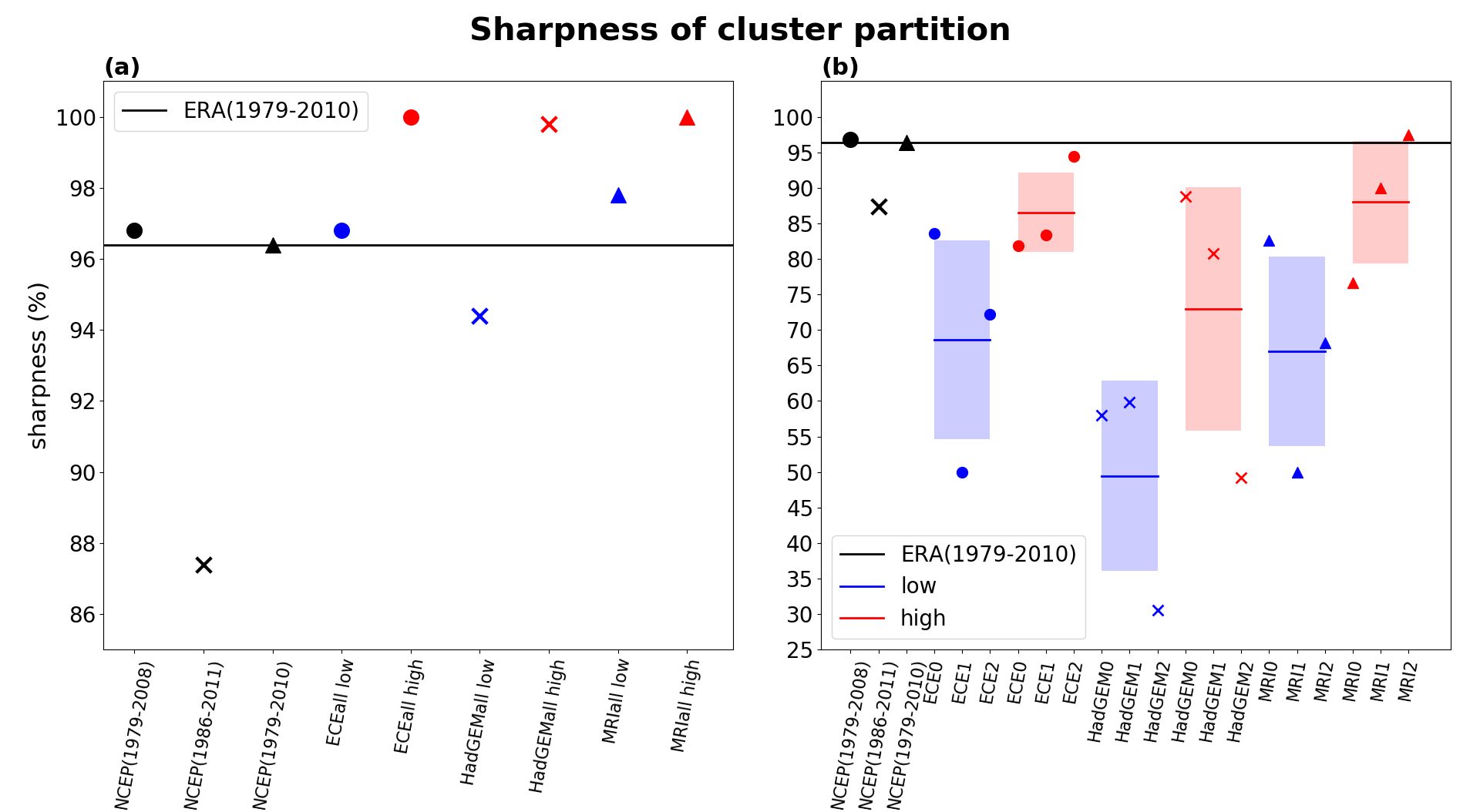}
\caption{The sharpness metric for re-analysis products and the three models. In (a), for NCEP re-analysis over the three different periods considered (black dots, crosses and triangles), along with that computed for all low/high EC-Earth resolution datasets concatenated (ECEall low/high, blue/red dots), all low/high HadGEM datasets concatenated (HadGEM low/high, blue/red crosses) and all low/high MRI datasets concatenated (MRIall low/high, blue/red triangles). In (b), sharpness for the same re-analysis data (black dots, crosses and triangles), and the sharpness metric for each individual ensemble member: EC-Earth low/high resolution (blue/red dots), HadGEM low/high resolution (blue/red crosses) and MRI low/high resolution (blue/red triangles). The mean and standard deviation are indicated by the horizontal coloured line and shading respectively. The horizontal black line in both plots shows the sharpness of ERA-Interim over the period 1979-2010.}
\label{fig:sigmetrics}
\end{figure}

Figure \ref{fig:sigmetrics} shows the impact of both increased resolution and increased ensemble size. In plot (a), the blue (red) dots/crosses/triangles associated with the ECEall/HadGEMall/MRIall labels are sharpness metrics for the relevant low (high) resolution model obtained after concatenating all three ensemble members, effectively tripling the sample size compared to the re-analysis datasets. The horizontal black line shows the sharpness of ERA-Interim over the period 1979-2010, while the black circles/crosses/triangles show the sharpness metrics for the NCEP re-analysis computed over the different time-periods covered by the three models. Note that the sharpness of NCEP during the period 1979-2010 matches that of ERA-Interim almost exactly, suggesting that this metric is well constrained.\ The two re-analysis datasets also produce nearly identical sharpness metrics when viewed over the two other time-periods. In plot (b), we show the sharpness metrics for each individual ensemble member of the three models at low resolution (blue dots/crosses/triangles) and high resolution (red dots/crosses/triangles). The blue/red lines in these triplets indicate the mean of the three points, and the shading encloses one standard deviation. 

The four sharpness metrics for re-analysis are almost identical (approximately 97\%), with the exception of that covering the period 1986-2011, where the metric is notably lower (approximately 87\%). As the time-period covered is shorter, it is possible that this is simply random sampling variability in a significantly clustered system. It is also possible that the extent to which regime dynamics drive the atmosphere is non-stationary, with the period 1979-1985 being particularly tightly clustered. However, since the concatenation of all three MRI experiments covering 1986-2011 have a sharpness close to 100\%, we will assume that the drop in sharpness seen with NCEP is sampling variability. Therefore, a difference in sharpness of up to 10\% might be expected by chance alone.

\begin{figure}[h]
\centering
\includegraphics[scale=0.5]{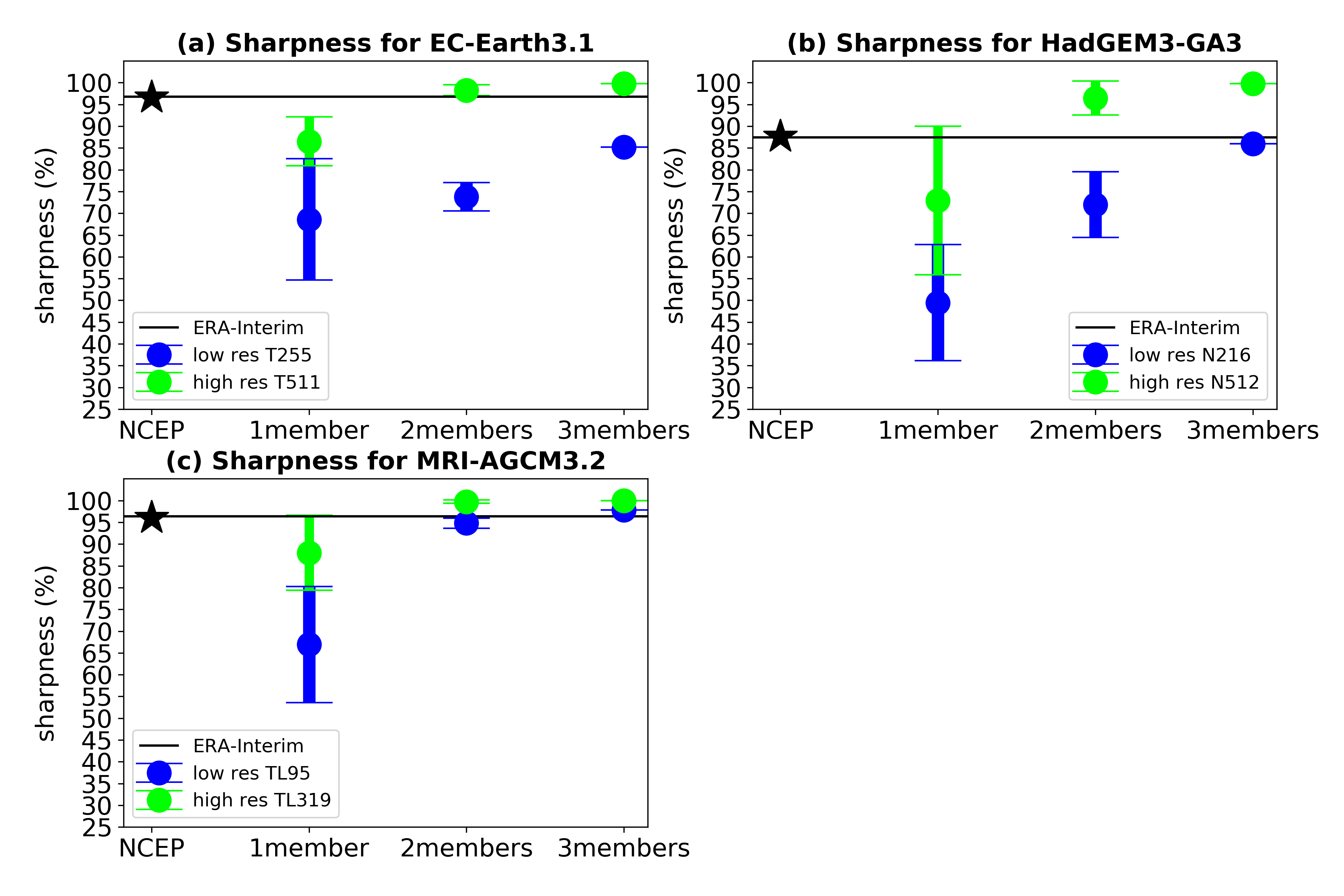}
\caption{Dependence of sharpness metric on number of ensemble members for (a) EC-Earth3.1, (b) HadGEM3-GA3, (c) MRI-AGCM3.2. In each, values for "1 member" is the average sharpness across the three low-resolution (respectively high-resolution) members, "2 members" the average sharpness when concatenating combinations of two ensemble members (over all such combinations) and "3 members" the sharpness obtained after concatenating all 3 members: low resolution in blue and high resolution in green. Error bars show one standard deviation around the mean. The horizontal black line shows the sharpness of ERA-Interim, and the black star shows the value of NCEP, over the relevant time-periods.}
\label{fig:sigperm}
\end{figure}

The impact of increased resolution is apparent for all three models, where sharpness increases whether looking at individual members or after concatenating all three. It can also be seen that increasing the sample size, by means of using more than 1 ensemble member, increases sharpness. This can be understood by noting that a model with weak regime structure will be more prone to producing clusters that cannot be robustly distinguished from random noise when using a small sample size. Increasing the sample size effectively serves to filter out noise in phase space. This can be examined further by evaluating the average sharpness metric obtained after concatenating two random members from each simulation. The results are shown in Figure \ref{fig:sigperm}, which summarizes the dependence of sharpness on ensemble size (i.e. sample size).\explain{Figure added to address an issue raised by reviewer 2.} It can be seen that the high-resolution models typically need twice the sample size of re-analysis data to achieve comparable regime structure, while for low resolution three times the sample size may still not suffice.

We note also (see Figure \ref{fig:sigmetrics}) that the increase in sharpness upon increasing the resolution is roughly twice as large as the maximum difference in sharpness between the two re-analysis products ERA-Interim and NCEP, with the former approximately 20\% and the latter at 10\%. This, combined with the fact that the increase is consistent across all three models, suggests that this improvement is statistically robust.

\subsection{Regime Locations}
\label{sec:regime_locations}

Table 1 shows the mean pattern correlation of the model clusters (across the three ensemble members) relative to ERA-Interim, with the error given as twice the standard deviation as computed across the three entries. High pattern correlation implies the cluster centroids of the model are in approximately the same location in phase space as re-analysis, and is therefore a measure of the regimes being in the correct location or not. In particular, the regime patterns will look similar to those in Figure \ref{fig:erapatterns} when this correlation is high. The contents of the table demonstrate that there is significant variability in this quantity, with no clear improvement across all three models upon increasing the resolution. 

 \begin{table}
 \label{tab}
 \caption{Mean pattern correlation of the regimes across the three simulations and errors (given as two standard deviations). Quantities labelled `All' are correlations of the model regimes obtained by concatenating all three simulations into one large dataset. Values from high resolution simulations are highlighted with bold to aid readability.}
 \centering
 \begin{tabular}{l c c c c}
 \hline
  & NAO+  & Blocking & Atlantic Ridge & NAO-  \\
 \hline
   ECE-Low  & $0.96 \,(\pm 0.02)$ & $0.84 \,(\pm 0.2)$ & $0.64 \,(\pm 0.63)$ & $0.87 \,(\pm 0.2)$   \\
   ECE-Hi  & $\textbf{0.81} \,(\pm 0.3)$  & $\textbf{0.78} \,(\pm 0.34)$ & $\textbf{0.49} \,(\pm 0.43)$ & $\textbf{0.90} \,(\pm 0.07)$   \\
   ECE-Low (all) & $0.96$ & $0.90$ & $0.85$ & $0.97$ \\
   ECE-Hi (all) & $\textbf{0.99}$ & $\textbf{0.96}$ & $\textbf{0.81}$ & $\textbf{0.90}$ \\ \hline
   HadGEM-Low & $0.96 \,(\pm 0.01)$ & $0.92 \,(\pm 0.03)$  & $0.93 \,(\pm 0.19)$ & $0.83 \,(\pm 0.38)$  \\
   HadGEM-Hi & $\textbf{0.93} \,(\pm 0.04)$  & $\textbf{0.89} \,(\pm 0.04)$ & $\textbf{0.91} \,(\pm 0.06)$ & $\textbf{0.80} \,(\pm 0.20)$  \\
   HadGem-Low (All) & $0.97$ & $0.96$ & $0.98$ & $0.99$ \\
   HadGem-Hi (All) & $\textbf{0.94}$ & $\textbf{0.89}$ & $\textbf{0.95}$ & $\textbf{0.79}$  \\ \hline
   MRI-Low & $0.74 \,(\pm 0.56)$ & $0.80 \,(\pm 0.13)$ & $0.71 \,\pm 0.54)$ & $0.88 \,(\pm 0.23)$  \\
   MRI-Hi  & $\textbf{0.89} \,(\pm 0.2)$ & $\textbf{0.75} \,(\pm 0.35)$  & $\textbf{0.74} \,(\pm 0.33)$ & $\textbf{0.95} \,(\pm 0.04)$  \\ 
   MRI-Low (All) & $0.94$ & $0.84$ & $0.96$ & $0.96$ \\
   MRI-Hi (All) & $\textbf{0.93}$ & $\textbf{0.78}$ & $\textbf{0.81}$ & $\textbf{0.96}$ \\ \hline
 \end{tabular}
 \end{table}

While for the Atlantic Ridge and NAO- regimes, the spread in the pattern correlation goes down when increasing resolution, for NAO+ and Blocking it frequently goes up. The mean correlation itself also shows no systematic improvement, and is sometimes degraded when increasing resolution. This can be seen more starkly in the `All' quantities in the table, which show the pattern correlations obtained from the model regimes after concatenating all three ensemble members. While we saw in the previous section that after combining the data in this way, the sharpness notably increased with resolution, the pattern correlation tends to slightly \emph{decrease}, with the average decrease across all models and regimes being approximately $-0.05$. The average change in pattern correlation across all un-concatenated experiments and all regimes, is approximately $-0.007$, with a standard deviation of $0.08$, indicating no significant change. Therefore the impact is at best neutral, and more likely a slight degradation.

These results suggest firstly that there remains considerable uncertainty in any model estimate of these quantities as diagnosed from even three ensemble members, and secondly that one cannot expect to see a consistent improvement in pattern correlation upon increasing resolution: the opposite may in fact be expected to happen. It is unclear to what extent sampling error is influencing the estimates: it is possible that with a much larger ensemble one would be able to spot more robust changes. Using all 10 EC-Earth members makes the uncertainty estimates smaller and much more uniform across the four regimes. However, the reduced uncertainty, of $\pm 0.16$ on average, still does not allow a statistically significant distinction between high and low resolution for all four regimes.

\subsection{Regime Persistence}
\label{sec:regime_persistence}

To assess the impact on persistence, we considered the seasonal persistence probabilities of the four regimes.\explain{The entire section has been shortened and the figure has been removed. This was to address a major issue raised by reviewer 2.} These are computed for a given season by estimating the probability that if the atmosphere is in a regime on day N, it will remain in the same regime on day N+1. This gives an indication of how persistent that regime tended to be in said season. For each model and each regime, distributions of these seasonal persistence probabilities were computed on the concatenation of the three ensemble members. These distributions were used to assess persistence.

All the models display biases in persistence statistics, but, with the exception of the blocking regime, we observed no systematic such bias, nor did we find a systematic improvement with resolution across all the models. For the blocking regime, all the models underestimate persistence, placing too much weight in the tail of short-lived events. This is consistent with previous studies showing that GCM's tend to underestimate persistent blocking (e.g. \citep{DAndrea1998}, \citep{Matsueda2009}, \citep{Anstey2013}, \citep{Masato2013}, \citep{Jung2012} and \citep{Davini2016}). The move to higher resolution results in all cases in the distribution shifting closer to re-analysis, implying that levels of persistence have increased for all three models. This is in agreement with the recent multi-model study conducted in \citep{Schiemann2017}, which also showed an improvement in blocking persistence upon increasing the resolution. Note that in all these papers atmospheric blocking was defined in a very different way, so our results are \emph{consistent} rather than equivalent. A figure showing this result can be found in the supporting material.

\section{Discussion and Conclusions}
\label{sec:conclusion}

All three models exhibit deficiencies in the geometric robustness of the regimes (measured with sharpness), the location of the regime centers (how similar the regime patterns are to those in re-analysis), and the persistence lifetime of regimes. Increasing the horizontal resolution leads to a notable improvement in the sharpness metric, suggesting that improvements in certain aspects of atmospheric dynamics are resulting in the atmosphere traversing phase space in a more tightly clustered fashion. Furthermore, the persistence statistics of the blocking regime did systematically improve, consistent with previous studies. However, the persistence statistics of the other three regimes, as well as the spatial pattern of all four regimes (including blocking) did not systematically improve, with some of the regimes deviating even further from re-analysis in some of the high-resolution simulations. This suggests that, while the increased horizontal resolution in these models is helping some crucial processes responsible for clustering behaviour, it is not having any notable effect on other regime biases.

Several studies (e.g. \citep{Davini2016}, \citep{Scaife2010}, \citep{Hinton2009}, \citep{Woollings2010}, \citep{Doblas-Reyes1998}) have implicated biases in the climatological mean state of models to errors in blocking statistics, and \citep{Masato2009} demonstrated the importance of the mean flow in determining the precise locations of blocking events. Given the synoptic scale nature of the Euro-Atlantic regimes, it is plausible that mean state biases may be important not just for the blocking regime, but for all four of the North Atlantic regimes, and these biases are not always systematically improved upon increasing horizontal resolution. Similarly, interannual variability due to global teleconnections such as ENSO, known to influence the NAO (see e.g. \citep{Li2012}), may change in non-obvious ways with resolution. Thus while certain aspects of the atmospheric dynamics may be improving from the increased resolution, leading to more pronounced clustering, the exact locations of these clusters may still be subject to errors due to other biases. This may be exacerbated in situations where the higher resolution model has not been tuned to achieve as realistic a mean state as the lower resolution version (as was the case with the models considered here). A detailed examination of this speculation is, however, beyond the scope of the present study.

The above discussion becomes more pertinent when observing that horizontal resolution cannot by itself account for regime skill. Indeed, Figure \ref{fig:sigmetrics} shows that the low resolution (180km) MRI simulations perform almost as well as the significantly higher resolution (25km) HadGEM simulations.\explain{This paragraph added in response to an issue raised by reviewer 2.} This implies that other aspects of the model, such as the model mean or small-scale variability related to differing physics parametrisations, are likely equally important for producing robust regimes. Interestingly, MRI also has the coarsest vertical resolution of the three models, seemingly ruling this out as an important factor.

Our results also show that the sampling variability in both the sharpness and spatial correlation metrics is large for both high and low resolution experiments. This may be due to excessive noise in the model atmosphere, or the weak regime structure allowing the model to populate phase space too liberally: in a potential well picture of the regimes, if the wells are too shallow, the model atmosphere will not stay trapped for longer periods of time, leading to less tightly clustered data. Either way, increasing the sample size by concatenating data from multiple ensemble members can help filter out this noise. We find that the low-resolution models appear to need three times as much data to detect a comparable regime structure as that found in re-analysis data, while the high-resolution models, by improving the regime structure, need only twice as much data. This implies that multiple ensemble members, \added{or a sufficiently long simulation period}\explain{Added in response to an issue by reviewer 2.}, are crucial for a statistically meaningful assessment of regime metrics, with changes due to resolution or other model upgrades being potentially completely invisible due to random noise.

%

\acknowledgments

Free accessibility of the EC-Earth SPHINX data to the climate user community is granted through a dedicated THREDDS Web Server hosted by CINECA (https://sphinx.hpc. cineca.it/thredds/sphinx.html), where it is possible to browse and directly download any of the Climate SPHINX data used in the present work. Further details on the data accessibility and on the Climate SPHINX project itself are available on the official website of the project (http://www.to.isac.cnr.it/ sphinx/). The SPHINX data was generated with computing resources provided by CINECA and LRZ in the framework of Climate SPHINX and Climate SPHINX reloaded PRACE projects. 

The MRI model integrations were performed using the Earth Simulator under the Development of Basic Technology for Risk Information on Climate Change of the SOUSEI Program of the Ministry of Education, Culture, Sports, Science and Technology (MEXT) of Japan. The model data was submitted as part of CMIP5 and is available on the CMIP5 data archive: https://cmip.llnl.gov/cmip5.

The UPSCALE data set was created by P. L. Vidale, M. Roberts, M. Mizielinski, J. Strachan, M.E. Demory and R. Schiemann using the HadGEM3 model, with support from NERC and the Met Office and the PRACE Research Infrastructure resource HERMIT, based in Germany at HLSR. We thank the large team of model developers, infrastructure experts, and all the other essential components required to conduct such a large-scale simulation campaign, in particular the PRACE infrastructure and the Stuttgart HLRS supercomputing center, as well as the STFC CEDA service for data storage and analysis using the JASMIN platform, where the data is publicly available. This work was supported by the Joint U.K. DECC/Defra Met Office Hadley Centre Climate Programme (GA01101). P. L. Vidale acknowledges the National Centre for Atmospheric Science Climate directorate (NCAS-Climate) (Contract R8/H12/83/001) for the High Resolution Climate Modelling (HRCM) program.

We acknowledge the contribution of the C3S 34a Lot 2 Copernicus Climate Change Service project, funded by the European Union, to the development of software tools used in this work. The clustering algorithm used to process the geopotential height fields, written with this support, is publicly available at https://github.com/IreMav/WRtool. The scripts used to plot the processed data is publicly available at https://github.com/KristianJS/regimes-grl. We thank Federico Fabiano (ISAC-CNR) for his contribution to the clustering software, and for doing some supporting computations involving all 10 EC-Earth ensemble members.

Jost von Hardenberg acknowledges support from the European Union’s Horizon 2020 research and innovation programme under grant agreement 641816 (CRESCENDO). Susanna Corti, Irene Mavilia and Kristian Strommen acknowledge support by the PRIMAVERA project, funded by the European Commission under grant agreement 641727 of the Horizon 2020 research program.


%
%
%
%
%
%



\begin{thebibliography}{29}
\providecommand{\natexlab}[1]{#1}
\expandafter\ifx\csname urlstyle\endcsname\relax
  \providecommand{\doi}[1]{doi:\discretionary{}{}{}#1}\else
  \providecommand{\doi}{doi:\discretionary{}{}{}\begingroup
  \urlstyle{rm}\Url}\fi



\bibitem[{\textit{Anstey et~al.}(2013)\textit{Anstey, Davini, Gray, Woollings,
  Butchart, Cagnazzo, Christiansen, Hardiman, Osprey, and Yang}}]{Anstey2013}
Anstey, J.~A., P.~Davini, L.~J. Gray, T.~J. Woollings, N.~Butchart,
  C.~Cagnazzo, B.~Christiansen, S.~C. Hardiman, S.~M. Osprey, and S.~Yang
  (2013), {Multi-model analysis of Northern Hemisphere winter blocking: Model
  biases and the role of resolution}, \textit{Journal of Geophysical Research
  Atmospheres}, \textit{118}(10), 3956--3971, \doi{10.1002/jgrd.50231}.

\bibitem[{\textit{Corti et~al.}(1999)\textit{Corti, Molteni, and
  Palmer}}]{Corti1999}
Corti, S., F.~Molteni, and T.~N. Palmer (1999), {Signature of recent climate
  change in frequencies of natural atmospheric circulation regimes},
  \textit{Nature}, \textit{398}(6730), 799--802, \doi{10.1038/19745}.

\bibitem[{\textit{D'Andrea et~al.}(1998)\textit{D'Andrea, Tibaldi, Blackburn,
  Boer, D{\'{e}}qu{\'{e}}, Dix, Dugas, Ferranti, Iwasaki, Kitoh, Pope, Randall,
  Roeckner, Straus, Stern, {Van Den Dool}, and Williamson}}]{DAndrea1998}
D'Andrea, F., S.~Tibaldi, M.~Blackburn, G.~Boer, M.~D{\'{e}}qu{\'{e}}, M.~R.
  Dix, B.~Dugas, L.~Ferranti, T.~Iwasaki, A.~Kitoh, V.~Pope, D.~Randall,
  E.~Roeckner, D.~Straus, W.~Stern, H.~{Van Den Dool}, and D.~Williamson
  (1998), {Northern Hemisphere atmospheric blocking as simulated by 15
  atmospheric general circulation models in the period 1979-1988},
  \textit{Climate Dynamics}, \textit{14}(6), 385--407,
  \doi{10.1007/s003820050230}.

\bibitem[{\textit{Davini and D'Andrea}(2016)}]{Davini2016}
Davini, P., and F.~D'Andrea (2016), {Northern Hemisphere atmospheric blocking
  representation in global climate models: Twenty years of improvements?},
  \textit{Journal of Climate}, \textit{29}(24), 8823--8840,
  \doi{10.1175/JCLI-D-16-0242.1}.

\bibitem[{\textit{Davini et~al.}(2017)\textit{Davini, {Von Hardenberg}, Corti,
  Christensen, Juricke, Subramanian, Watson, Weisheimer, and
  Palmer}}]{Davini2017}
Davini, P., J.~{Von Hardenberg}, S.~Corti, H.~M. Christensen, S.~Juricke,
  A.~Subramanian, P.~A. Watson, A.~Weisheimer, and T.~N. Palmer (2017),
  {Climate SPHINX: Evaluating the impact of resolution and stochastic physics
  parameterisations in the EC-Earth global climate model},
  \textit{Geoscientific Model Development}, \textit{10}(3), 1383--1402,
  \doi{10.5194/gmd-10-1383-2017}.

\bibitem[{\textit{Dawson and Palmer}(2015)}]{Dawson2015}
Dawson, A., and T.~N. Palmer (2015), {Simulating weather regimes: impact of
  model resolution and stochastic parameterization}, \textit{Climate Dynamics},
  \textit{44}(7-8), 2177--2193, \doi{10.1007/s00382-014-2238-x}.

\bibitem[{\textit{Dawson et~al.}(2012)\textit{Dawson, Palmer, and
  Corti}}]{Dawson2012}
Dawson, A., T.~N. Palmer, and S.~Corti (2012), {Simulating regime structures in
  weather and climate prediction models}, \textit{Geophysical Research
  Letters}, \textit{39}(21), \doi{10.1029/2012GL053284}.

\bibitem[{\textit{Dee et~al.}(2011)\textit{Dee, Uppala, Simmons, Berrisford,
  Poli, Kobayashi, Andrae, Balmaseda, Balsamo, Bauer, Bechtold, Beljaars,
  van~de Berg, Bidlot, Bormann, Delsol, Dragani, Fuentes, Geer, Haimberger,
  Healy, Hersbach, H{\'{o}}lm, Isaksen, K{\aa}llberg, K{\"{o}}hler, Matricardi,
  Mcnally, Monge-Sanz, Morcrette, Park, Peubey, de~Rosnay, Tavolato,
  Th{\'{e}}paut, and Vitart}}]{Dee2011}
Dee, D.~P., S.~M. Uppala, A.~J. Simmons, P.~Berrisford, P.~Poli, S.~Kobayashi,
  U.~Andrae, M.~A. Balmaseda, G.~Balsamo, P.~Bauer, P.~Bechtold, A.~C.
  Beljaars, L.~van~de Berg, J.~Bidlot, N.~Bormann, C.~Delsol, R.~Dragani,
  M.~Fuentes, A.~J. Geer, L.~Haimberger, S.~B. Healy, H.~Hersbach, E.~V.
  H{\'{o}}lm, L.~Isaksen, P.~K{\aa}llberg, M.~K{\"{o}}hler, M.~Matricardi,
  A.~P. Mcnally, B.~M. Monge-Sanz, J.~J. Morcrette, B.~K. Park, C.~Peubey,
  P.~de~Rosnay, C.~Tavolato, J.~N. Th{\'{e}}paut, and F.~Vitart (2011), {The
  ERA-Interim reanalysis: Configuration and performance of the data
  assimilation system}, \textit{Quarterly Journal of the Royal Meteorological
  Society}, \textit{137}(656), 553--597, \doi{10.1002/qj.828}.

\bibitem[{\textit{Doblas-Reyes et~al.}(1998)\textit{Doblas-Reyes,
  D{\'{e}}qu{\'{e}}, Valero, and Stephenson}}]{Doblas-Reyes1998}
Doblas-Reyes, F.~J., M.~D{\'{e}}qu{\'{e}}, F.~Valero, and D.~B. Stephenson
  (1998), {North Atlantic wintertime intraseasonal variability and its
  sensitivity to GCM horizontal resolution}, \textit{Tellus, Series A: Dynamic
  Meteorology and Oceanography}, \textit{50}(5), 573--595,
  \doi{10.3402/tellusa.v50i5.14560}.

\bibitem[{\textit{Ferranti et~al.}(2015)\textit{Ferranti, Corti, and
  Janousek}}]{Ferranti2015}
Ferranti, L., S.~Corti, and M.~Janousek (2015), {Flow-dependent verification of
  the ECMWF ensemble over the Euro-Atlantic sector}, \textit{Quarterly Journal
  of the Royal Meteorological Society}, \textit{141}(688), 916--924,
  \doi{10.1002/qj.2411}.

\bibitem[{\textit{Frame et~al.}(2013)\textit{Frame, Methven, Gray, and
  Ambaum}}]{Frame2013}
Frame, T.~H., J.~Methven, S.~L. Gray, and M.~H. Ambaum (2013), {Flow-dependent
  predictability of the North Atlantic jet}, \textit{Geophysical Research
  Letters}, \textit{40}(10), 2411--2416, \doi{10.1002/grl.50454}.

\bibitem[{\textit{Franzke et~al.}(2011)\textit{Franzke, Woollings, and
  Martius}}]{Franzke2011}
Franzke, C., T.~Woollings, and O.~Martius (2011), {Persistent Circulation
  Regimes and Preferred Regime Transitions in the North Atlantic},
  \textit{Journal of the Atmospheric Sciences}, \textit{68}(12), 2809--2825,
  \doi{10.1175/JAS-D-11-046.1}.

\bibitem[{\textit{Hannachi et~al.}(2017)\textit{Hannachi, Straus, Franzke,
  Corti, and Woollings}}]{Hannachi2017}
Hannachi, A., D.~M. Straus, C.~L. Franzke, S.~Corti, and T.~Woollings (2017),
  {Low-frequency nonlinearity and regime behavior in the Northern Hemisphere
  extratropical atmosphere}, \doi{10.1002/2015RG000509}.

\bibitem[{\textit{Hazeleger et~al.}(2012)\textit{Hazeleger, Wang, Severijns,
  Ştefǎnescu, Bintanja, Sterl, Wyser, Semmler, Yang, van~den Hurk, van Noije,
  van~der Linden, and van~der Wiel}}]{Hazeleger2012}
Hazeleger, W., X.~Wang, C.~Severijns, S.~Ştefǎnescu, R.~Bintanja, A.~Sterl,
  K.~Wyser, T.~Semmler, S.~Yang, B.~van~den Hurk, T.~van Noije, E.~van~der
  Linden, and K.~van~der Wiel (2012), {EC-Earth V2.2: Description and
  validation of a new seamless earth system prediction model}, \textit{Climate
  Dynamics}, \textit{39}(11), 2611--2629, \doi{10.1007/s00382-011-1228-5}.

\bibitem[{\textit{Hinton et~al.}(2009)\textit{Hinton, Hoskins, and
  Martin}}]{Hinton2009}
Hinton, T.~J., B.~J. Hoskins, and G.~M. Martin (2009), {The influence of
  tropical sea surface temperatures and precipitation on north Pacific
  atmospheric blocking}, \textit{Climate Dynamics}, \textit{33}(4), 549--563,
  \doi{10.1007/s00382-009-0542-7}.

\bibitem[{\textit{Jung et~al.}(2012)\textit{Jung, Miller, Palmer, Towers, Wedi,
  Achuthavarier, Adams, Altshuler, Cash, Kinter, Marx, Stan, and
  Hodges}}]{Jung2012}
Jung, T., M.~J. Miller, T.~N. Palmer, P.~Towers, N.~Wedi, D.~Achuthavarier,
  J.~M. Adams, E.~L. Altshuler, B.~A. Cash, J.~L. Kinter, L.~Marx, C.~Stan, and
  K.~I. Hodges (2012), {High-resolution global climate simulations with the
  ECMWF model in project athena: Experimental design, model climate, and
  seasonal forecast skill}, \textit{Journal of Climate}, \textit{25}(9),
  3155--3172, \doi{10.1175/JCLI-D-11-00265.1}.

\bibitem[{\textit{Kalnay et~al.}(1996)\textit{Kalnay, Kanamitsu, Kistler,
  Collins, Deaven, Gandin, Iredell, Saha, White, Woollen, Zhu, Chelliah,
  Ebisuzaki, Higgins, Janowiak, Mo, Ropelewski, Wang, Leetmaa, Reynolds, Jenne,
  and Joseph}}]{Kalnay1996}
Kalnay, E., M.~Kanamitsu, R.~Kistler, W.~Collins, D.~Deaven, L.~Gandin,
  M.~Iredell, S.~Saha, G.~White, J.~Woollen, Y.~Zhu, M.~Chelliah, W.~Ebisuzaki,
  W.~Higgins, J.~Janowiak, K.~C. Mo, C.~Ropelewski, J.~Wang, A.~Leetmaa,
  R.~Reynolds, R.~Jenne, and D.~Joseph (1996), {The NCEP/NCAR 40-year
  reanalysis project}, \textit{Bulletin of the American Meteorological
  Society}, \textit{77}(3), 437--471,
  \doi{10.1175/1520-0477(1996)077<0437:TNYRP>2.0.CO;2}.

\bibitem[{Kob(2015)}]{Kobayashi2015}
 (2015), {The JRA-55 reanalysis: General specifications and basic
  characteristics}, \textit{Journal of the Meteorological Society of Japan.
  Ser. II}, \textit{93}(1), 5--48, \doi{10.2151/jmsj.2015-001}.

\bibitem[{\textit{Li and Lau}(2012)}]{Li2012}
Li, Y., and N.~C. Lau (2012), {Impact of ENSO on the atmospheric variability
  over the North Atlantic in late Winter-Role of transient eddies},
  \textit{Journal of Climate}, \textit{25}(1), 320--342,
  \doi{10.1175/JCLI-D-11-00037.1}.

\bibitem[{\textit{Masato et~al.}(2009)\textit{Masato, Hoskins, and
  Woollings}}]{Masato2009}
Masato, G., B.~J. Hoskins, and T.~J. Woollings (2009), {Can the Frequency of
  Blocking Be Described by a Red Noise Process?}, \textit{Journal of the
  Atmospheric Sciences}, \textit{66}(7), 2143--2149,
  \doi{10.1175/2008JAS2907.1}.

\bibitem[{\textit{Masato et~al.}(2013)\textit{Masato, Hoskins, and
  Woollings}}]{Masato2013}
Masato, G., B.~J. Hoskins, and T.~Woollings (2013), {Winter and Summer Northern
  Hemisphere Blocking in CMIP5 Models}, \textit{Journal of Climate},
  \textit{26}(18), 7044--7059, \doi{10.1175/JCLI-D-12-00466.1}.

\bibitem[{\textit{Matsueda and Palmer}(2018)}]{Matsueda2018}
Matsueda, M., and T.~Palmer (2018), {Estimates of flow-dependent predictability
  of wintertime Euro-Atlantic weather regimes in medium-range forecasts},
  \textit{Quarterly Journal of the Royal Meteorological Society},
  \doi{10.1002/qj.3265}.

\bibitem[{\textit{Matsueda et~al.}(2009)\textit{Matsueda, Mizuta, and
  Kusunoki}}]{Matsueda2009}
Matsueda, M., R.~Mizuta, and S.~Kusunoki (2009), {Future change in wintertime
  atmospheric blocking simulated using a 20-km-mesh atmospheric global
  circulation model}, \textit{Journal of Geophysical Research Atmospheres},
  \textit{114}(12), \doi{10.1029/2009JD011919}.

\bibitem[{\textit{Michelangeli et~al.}(1995)\textit{Michelangeli, Vautard, and
  Legras}}]{Michelangeli1995}
Michelangeli, P.-A., R.~Vautard, and B.~Legras (1995), {Weather Regimes:
  Recurrence and Quasi Stationarity}, \textit{Journal of the Atmospheric
  Sciences}, \textit{52}(8), 1237--1256,
  \doi{10.1175/1520-0469(1995)052<1237:WRRAQS>2.0.CO;2}.

\bibitem[{\textit{Mizielinski et~al.}(2014)\textit{Mizielinski, Roberts,
  Vidale, Schiemann, Demory, Strachan, Edwards, Stephens, Lawrence, Pritchard,
  Chiu, Iwi, Churchill, {Del Cano Novales}, Kettleborough, Roseblade, Selwood,
  Foster, Glover, and Malcolm}}]{Mizielinski2014}
Mizielinski, M.~S., M.~J. Roberts, P.~L. Vidale, R.~Schiemann, M.~E. Demory,
  J.~Strachan, T.~Edwards, A.~Stephens, B.~N. Lawrence, M.~Pritchard, P.~Chiu,
  A.~Iwi, J.~Churchill, C.~{Del Cano Novales}, J.~Kettleborough, W.~Roseblade,
  P.~Selwood, M.~Foster, M.~Glover, and A.~Malcolm (2014), {High-resolution
  global climate modelling: The UPSCALE project, a large-simulation campaign},
  \textit{Geoscientific Model Development}, \doi{10.5194/gmd-7-1629-2014}.



\bibitem[{\textit{Mizuta et~al.}(2012)\textit{Mizuta, Yoshimura, Murakami, Matsueda, Endo, Ose, Kamiguchi, Hosaka, Sugi, Yukimoto, Kusunoki and Kitoh}}]{Mizuta2012}
Mizuta, R.~H., H.~Yoshimura, M.~Murakami, H.~Matsueda, T.~Endo, K.~Ose, M.~Kamiguchi, M.~Hosaka, M.~Sugi, S.~Yukimoto, S.~Kusunoki, and A.~Kitoh (2012), {Climate simulations using MRI-AGCM3.2 with 20-km grid}, \textit{J. Meteor. Soc. Japan}, \textit{90A}, 233--258.


\bibitem[{\textit{Palmer}(1999)}]{Palmer1999}
Palmer, T.~N. (1999), {A nonlinear dynamical perspective on climate
  prediction}, \textit{Journal of Climate}, \textit{12}(2), 575--591,
  \doi{10.1175/1520-0442(1999)012<0575:ANDPOC>2.0.CO;2}.

\bibitem[{\textit{Scaife et~al.}(2010)\textit{Scaife, Woollings, Knight,
  Martin, and Hinton}}]{Scaife2010}
Scaife, A.~A., T.~Woollings, J.~Knight, G.~Martin, and T.~Hinton (2010),
  {Atmospheric blocking and mean biases in climate models}, \textit{Journal of
  Climate}, \textit{23}(23), 6143--6152, \doi{10.1175/2010JCLI3728.1}.

\bibitem[{\textit{Schiemann et~al.}(2017)\textit{Schiemann, Demory, Shaffrey,
  Strachana, Vidale, Mizielinski, Roberts, Matsueda, Wehner, Jung, and
  Jung}}]{Schiemann2017}
Schiemann, R., M.~E. Demory, L.~C. Shaffrey, J.~Strachana, P.~L. Vidale, M.~S.
  Mizielinski, M.~J. Roberts, M.~Matsueda, M.~F. Wehner, T.~Jung, and T.~Jung
  (2017), {The resolution sensitivity of Northern Hemisphere blocking in four
  25-km atmospheric global circulation models}, \textit{Journal of Climate},
  \textit{30}(1), 337--358, \doi{10.1175/JCLI-D-16-0100.1}.

\bibitem[{\textit{Straus}(2010)}]{Straus2010}
Straus, D.~M. (2010), {Synoptic-Eddy Feedbacks and Circulation Regime
  Analysis}, \textit{Monthly Weather Review}, \textit{138}(11), 4026--4034,
  \doi{10.1175/2010MWR3333.1}.

\bibitem[{\textit{Straus et~al.}(2007)\textit{Straus, Corti, and
  Molteni}}]{Straus2007}
Straus, D.~M., S.~Corti, and F.~Molteni (2007), {Circulation regimes: Chaotic
  variability versus SST-forced predictability}, \textit{Journal of Climate},
  \textit{20}(10), 2251--2272, \doi{10.1175/JCLI4070.1}.

\bibitem[{\textit{Walters et~al.}(2011)\textit{Walters, Best, Bushell, Copsey,
  Edwards, Falloon, Harris, Lock, Manners, Morcrette, Roberts, Stratton,
  Webster, Wilkinson, Willett, Boutle, Earnshaw, Hill, MacLachlan, Martin,
  Moufouma-Okia, Palmer, Petch, Rooney, Scaife, and Williams}}]{Walters2011}
Walters, D.~N., M.~J. Best, A.~C. Bushell, D.~Copsey, J.~M. Edwards, P.~D.
  Falloon, C.~M. Harris, A.~P. Lock, J.~C. Manners, C.~J. Morcrette, M.~J.
  Roberts, R.~A. Stratton, S.~Webster, J.~M. Wilkinson, M.~R. Willett, I.~A.
  Boutle, P.~D. Earnshaw, P.~G. Hill, C.~MacLachlan, G.~M. Martin,
  W.~Moufouma-Okia, M.~D. Palmer, J.~C. Petch, G.~G. Rooney, A.~A. Scaife, and
  K.~D. Williams (2011), {The Met Office Unified Model Global Atmosphere
  3.0/3.1 and JULES Global Land 3.0/3.1 configurations}, \textit{Geoscientific
  Model Development}, \textit{4}(4), 919--941, \doi{10.5194/gmd-4-919-2011}.

\bibitem[{\textit{Woollings et~al.}(2010{\natexlab{a}})\textit{Woollings,
  Hannachi, Hoskins, and Turner}}]{Woollings2010a}
Woollings, T., A.~Hannachi, B.~Hoskins, and A.~Turner (2010{\natexlab{a}}), {A
  regime view of the North Atlantic oscillation and its response to
  anthropogenic forcing}, \textit{Journal of Climate}, \textit{23}(6),
  1291--1307, \doi{10.1175/2009JCLI3087.1}.

\bibitem[{\textit{Woollings et~al.}(2010{\natexlab{b}})\textit{Woollings,
  Hannachi, and Hoskins}}]{Woollings2010b}
Woollings, T., A.~Hannachi, and B.~Hoskins (2010{\natexlab{b}}), {Variability
  of the North Atlantic eddy-driven jet stream}, \textit{Quarterly Journal of
  the Royal Meteorological Society}, \textit{136}(649), 856--868,
  \doi{10.1002/qj.625}.

\bibitem[{\textit{Woollings et~al.}(2010{\natexlab{c}})\textit{Woollings,
  Charlton-Perez, Ineson, Marshall, and Masato}}]{Woollings2010}
Woollings, T., A.~Charlton-Perez, S.~Ineson, A.~G. Marshall, and G.~Masato
  (2010{\natexlab{c}}), {Associations between stratospheric variability and
  tropospheric blocking}, \textit{Journal of Geophysical Research Atmospheres},
  \textit{115}(6), \doi{10.1029/2009JD012742}.

\end{thebibliography}




\listofchanges

\end{document}